# The First Historical Standstill of WW Ceti

**Mike Simonsen**
*AAVSO, 49 Bay State Road, Cambridge, MA 02138; mikesimonsen@aavso.org*

**Rod Stubbings**
*Tetoora Observatory, 2643 Warragul-Korumburra Road, Tetoora Road 3821, Victoria, Australia; stubbo@sympac.com.au*

**Abstract**    Z Cam dwarf novae are distinguished from other dwarf novae based on the appearance of so called "standstills" in their long-term optical light curves. It has been suggested previously that WW Cet might be a Z Cam type dwarf nova, but this classification was subsequently ruled out, based on its long-term light curve behavior. Forty years of historical data for WW Cet has shown no evidence of standstills. WW Cet is therefore classified as a UG type dwarf nova in the *General Catalogue of Variable Stars* (GCVS, Samus *et al.* 2007–2009) and the International Variable Star Index (VSX, Watson *et al.* 2006). Beginning in the 2010 observing season, WW Cet has been observed to be in a standstill, remaining more or less steady in the 12th magnitude range. Based on this first ever, historical standstill of WW Cet, we conclude that it is indeed a bona fide member of the Z Cam class of dwarf novae.

### 1. Introduction

U Geminorum-type (UG) variables, also called dwarf novae, are close binary systems consisting of a dwarf or subgiant K-M star that fills the volume of its inner Roche lobe. This donor star is losing mass to a white dwarf surrounded by an accretion disk. From time to time the system goes into outburst, brightening rapidly by several magnitudes. After several days to a month, or more, it returns to its original state.

These dwarf nova outbursts are thought to be caused by thermal instabilities in the disk. Gas accumulates in the disk until it heats up and becomes viscous. This increased viscosity causes it to migrate in toward the white dwarf, heating up even more, eventually causing an outburst.

Intervals between consecutive outbursts for a given star may vary, but every star is characterized by a characteristic mean value of these intervals. This mean cycle corresponds to the mean amplitude of the outbursts. Generally speaking, the longer the cycle, the greater the amplitude of the outbursts.

According to the characteristics of their light curves, UGs are further subdivided into three types: SS Cyg (UGSS), SU UMa (UGSU), and Z Cam (UGZ). UGSU variables are not relevant to this discussion.

UGSS increase in brightness by 2 to 6 magnitudes in *V* in 1 to 2 days. After several subsequent days, they return to their original brightness. The cycle times vary considerably, from 10 to hundreds of days.

UGZ also show cyclic outbursts, but differ from UGSS variables by the fact that sometimes after an outburst they do not return to their quiescent magnitude. Instead, they appear to get stuck, for months or even years, at a brightness of one to one and a half magnitudes fainter than outburst maximum. These episodes are known as standstills. Z Cam cycle times characteristically range from 10 to 40 days, and their outburst amplitudes are from 2 to 5 magnitudes in *V*, but standstills are the defining characteristic of the Z Cam stars.

**2. History**

Luyten (Liller 1962a) was the first to mention WW Cet as a variable star. He also suspected it was a cataclysmic variable (CV). Herbig (Liller 1962b) confirmed the CV nature of the star. The first to suggest WW Cet might belong to the Z Cam class of dwarf novae was Paczynski (1963). The system was catalogued as UGZ in the GCVS (Kukarkin *et al.* 1969).

Warner (1987) and Ringwald *et al.* (1996) concluded that since the long-term light curve behavior of WW Cet showed no evidence of standstills it was not a bona fide member of the Z Cam subclass. Both the GCVS (Samus *et al.* 2007–2009) and VSX currently list WW Cet as a UG.

**3. Characteristics**

A detailed inspection of the AAVSO data shows that WW Cet normally ranges from 16.0V at minimum to 10.5V in outburst. From August 1968 through the end of 2009 its cyclic pattern of outbursts and quiescent stages appears to be as a UGSS star (Figure 1). It has an average cycle time between outbursts of 31.2 days (Samus *et al.* 2007–2009).

The physical characteristics of the star are fairly well known. The orbital period is 0.17587 days (4.22 hours) and the masses of the primary and secondary are known to be 1.05 and 0.393 solar masses respectively. The orbital inclination is 48 degrees ($\pm$ 11), so WW Cet does not exhibit eclipses (Tappert *et al.* 1997).

It is the longish period, above the period gap, that has led some in the past to suspect WW Cet might be a UGZ. All confirmed UGZ have periods longer than 3 hours. However, this characteristic is not proof of membership in the Z Cam class.

**4. The 2010 standstill**

There is solid observational evidence in the AAVSO International Database that WW Cet has been in standstill for at least 78 days (Figure 2). The earliest observation of the 2010 standstill event was reported on September 10, 2010. WW

Cet was reported to be magnitude 12.8v. Through November 27, 2010 the star remained 12th magnitude, averaging ~12.4 with fluctuations from 12.9V to 12.0v (Figure 3).

This is 3 to 4 magnitudes in *V* above its typical minimum value and ~1.5 magnitudes in *V* below the average maximum value, almost textbook agreement with the definition of a UGZ standstill. Add to this the fact that its outburst amplitude, cycle time and orbital period all conform to the characteristics of UGZ classification and the evidence for WW Cet as a bona fide Z Cam dwarf nova is overwhelming.

**5. Conclusion**

We conclude that the behavior of WW Cet in the 2010 observing season represents the first ever recorded standstill of this dwarf novae. Combined with the known characteristics of amplitude, cycle time and orbital period, this standstill confirms WW Cet is a Z Cam dwarf nova.

**6. Acknowledgements**

We acknowledge with thanks the variable star observations from the AAVSO International Database contributed by observers worldwide and used in this research.

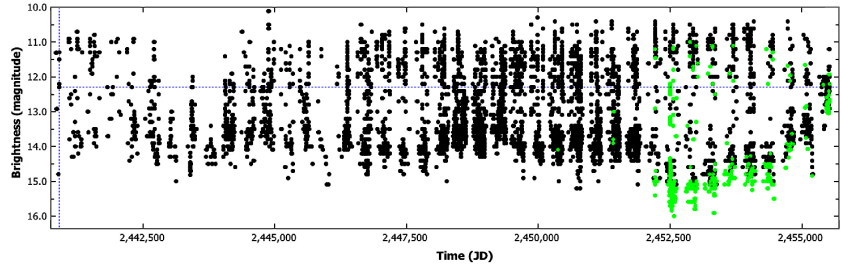

Figure 1. Light curve for WW Cet (1970–2009) using long-term AAVSO data, and demonstrating the UGSS nature of the outburst cycle. Visual observations (dark points), and Johnson *V* observations (lighter points) are shown.

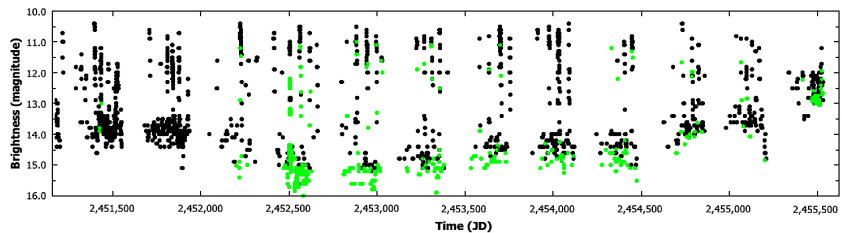

Figure 2. Light curve for WW Cet using AAVSO data from January 1, 1999 to November 27, 2010. The normal UGSS cyclic behavior and the typical range of variation from the prior eleven observing seasons are in stark contrast to the last three months of data, including the 78 days where WW Cet has remained in standstill between 12th and 13th magnitude. Visual observations (dark points), and Johnson *V* observations (lighter points) are shown.

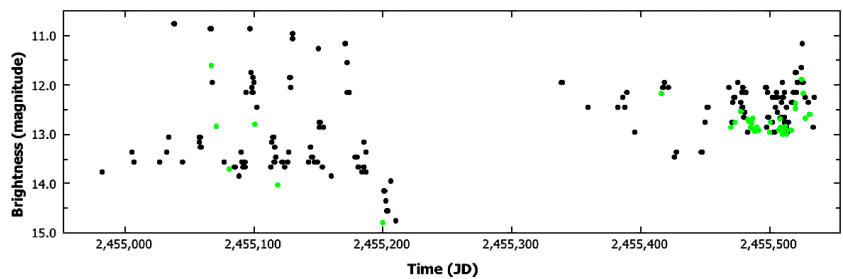

Figure 3. Light curve for WW Cet using AAVSO data from the 2009 observing season (left) showing the UGSS-like outburst cycle, and data from the 2010 observing season showing the first recorded standstill of WW Cet. Visual observations (dark points), and Johnson *V* observations (lighter points) are shown.